\documentclass{aa}

\usepackage{graphicx}
\usepackage{txfonts}
 \newcommand{\funitsS}{10$^{-17}$ erg s$^{-1}$ cm$^{-2}$ arcsec$^{-2}$}
 \newcommand{\vel}{km s$^{-1}$\ }

\begin{document}
%
%   \title{R3D, a package for reducing fiber-based Integral Field Spectroscopy
%   data}
   \title{Integral Field Spectroscopy of two radio galaxies at z$\sim$2.3}
   \author{S.F.S\'anchez\thanks{Based on observations collected at the
   Centro Astron\'mico Hispano Alem\'an (CAHA) at Calar Alto, operated jointly
   by the Max-Planck Institut f\"ur Astronomie and the Instituto de Astrof\'\i
   sica de Andaluc\'\i a (CSIC).}
          \inst{1}
%          \and
%          M.Villar-Mart{\'{\i}}n\inst{2}
          \and
          A.Humphrey{$^2$}
%	  \and
%	  R.F.Peletier{5}
          }

 \offprints{S.F.S\'anchez, sanchez@caha.es}
\institute{Centro Astron\'omico Hispano Alem\'an, Calar Alto, (CSIC-MPG),
  C/Jes\'us Durb\'an Rem\'on 2-2, E-04004 Almeria, Spain (sanchez@caha.es).
%  \and
%  Instituto de Astrof\'\i sica de Andaluc\'\i a, Camino Bajo de Huetor, S/N,
%  3004 Gradana, Spain.
  \and
  Korea Astromomy and Space Science Institute, 61-1 Hwaam-dong, Yuseong-gu, Daejeon, 305-348, Republic of Korea
%  \and
%  Kapteyn Astronomical Institute, PO Box 800, 9700 AV Groningen, The Netherlands
}
\date{;}

\abstract{}{In this article we study the morphology, kinematics and
ionization properties of the giant ionized gas nebulae surrounding two high
redshift radio galaxies, 4C40.36 ($z=$2.27) and 4C48.48 ($z=$2.34).}{Integral
Field Spectroscopy observations were taken using the PPAK bundle of the PMAS
spectrograph, mounted on the 3.5m on the Calar Alto Observatory, in order to
cover a field-of-view of 64$\arcsec\times$72$\arcsec$ centered in each radio
galaxy. The observations spanned over 5 nights, using two different spectral
resolutions (with FWHM$\sim$4\AA\ and $\sim$8\AA\ respectively), covering the
optical wavelength range from $\sim$3700\AA\ to $\sim$7100\AA, which
corresponds to the rest-frame ultraviolet range from $\sim$1100\AA\ to
$\sim$2000\AA\ . Various emission
lines are detected within this wavelength range, including Ly$\alpha$
(1216\AA), NV (1240\AA), CIV (1549\AA), HeII (1640\AA), OIII] (1663\AA) and
CIII] (1909\AA). The dataset was used to derive the spatial distribution of
the flux intensity of each of these lines and the gas kinematics. The
properties of the emission lines in the nuclear regions were studied in
detail.}{In agreement with previous studies, we find that both objects are
embedded in a large ionized gas nebula, where Ly$\alpha$ emission 
is extended across
$\sim$100 kpc or more. The CIV and HeII emission lines are also spatially
extended. The nebulae are generally aligned with the radio axis, although we
detect emission far from it. In  4C+48.48, there is
a band of low Ly$\alpha$/CIV running perpendicular to the radio axis, at the
location of the active nucleus. This feature might be the observational
signature of an edge-on disk of neutral gas. The kinematics of both nebulae
are inconsistent with stable rotation, although they are not inconsistent with
infall or outflow.}{}

%\abstract{}{ADASDAD}


 \maketitle

\section{Introduction}

High redshift radio galaxies (HzRGs) are often surrounded by giant Ly$\alpha$
nebulae which sometimes extend for more than 100 kpc (e.g., McCarthy et
al. 1990a; Reuland et al. 2003; Villar-Mart{\'{\i}}n et al. 2003;
Villar-Mart{\'{\i}}n et al. 2007b) and sometimes beyond the radio structure
(e.g., Eales et al. 1993; Kurk et al. 2002; Maxfield et al. 2002). Giant
Ly$\alpha$ nebulae have also been found associated with other classes of
high-$z$ object, both active and non active ones, radio loud and radio quiet
(e.g., Christensen et al. 2006; Steidel et al. 2001). They are frequently
aligned with the radio axis (McCarthy et al. 1995), showing an irregular
morphology (e.g., Reuland et al. 2003).

In the rest-frame ultraviolet (UV) wavelength range, the giant nebulae
associated with HzRGs show a variety of strong emission lines, with Ly$\alpha$
usually being the brightest, followed by CIV$\lambda$1550, HeII$\lambda$1640
and CIII]$\lambda1909$ (hereafter CIV, HeII and CIII]).  A multitude of weaker
lines have also been detected in deep spectra (e.g. Vernet et al. 2001).
Based on comparisons between emission line ratios and various ionization
models, the line emitting gas is substantially enriched with metals, and is
ionized predomonantly by the hard radiation field of the active nucleus (e.g.,
Vernet et al. 2001; Humphrey et al. 2008). In a few extreme cases, however,
young stars or shocks may also make a significant contribution to the
ionization of this gas (e.g. Villar-Mart{\'{\i}}n et al. 2007a; Maxfield et
al. 2002).  The nebulae have typical masses of $\sim$10$^{\rm 9-10}$
M$_\odot$, Ly$\alpha$ luminosities of the order of $\sim$10$^{\rm 43-44}$ erg
s$^{-1}$ and electron densities of a few to several hundred cm$^{-3}$
(e.g.,McCarthy 1993; Reuland et al. 2003).

These nebulae often show high velocity dispersions (FWHM$>$1000 \vel)
associated with, and confined by, the radio structures (e.g.,
Villar-Mart{\'{\i}}n et al. 2003). It is likely that these large velocity
dispersions are a consequence of interactions between the radio structure and
the ambient interstellar medium, which are thought to result in outflows
(e.g., Villar-Mart{\'{\i}}n et al. 2003; Humphrey et al. 2006; van Ojik et
al. 1997).

In addition, the giant nebulae associated with HzRGs also contain gas with
significantly lower velocity dispersions, which shows no clear relationship
with the radio source, and in many cases is observed to extend far beyond it
(i.e., $\sim$500 \vel: Villar-Mart{\'{\i}}n et al. 2002; 2003). These nebulae
seem to be present too, although not in all the cases, in another kind of
AGNs, like radio-quiet QSOs (e.g., Christensen et al. 2006, Husemann et
al. 2008).  Humphrey et al. (2007) have recently proposed that this {\it
quiescent} gas is in infall towards the central regions of the host
galaxy. The importance of this gas in feeding the AGN activity, as a reservoir
of gas for galaxy assembling at high-redshift (e.g., di Matteo et al. 2005),
or its role in the AGN feedback process is still unclear (e.g., Kang et
al. 2007; Best 2007).

Most previous studies of the giant ionized nebulae around HzRGs have been
undertaken using long-slit spectroscopy, usually with the slit aligned with
the radio structure (e.g. Villar-Mart{\'{\i}}n et al. 2003).  A few other
studies have made use of narrow band images of Ly$\alpha$ (e.g. Reuland et
al. 2003).  While both of these types of observation have resulted in a
significant expansion in our understanding of the giant nebulae, they were
limited by a lack of spatial information or a lack of spectral information,
respectively.  Integral field spectroscopy (IFS) has a clear advantage over
both of these methods, in that spectral and 2-dimensional spatial information
can be obtained simultaneously.  For this reason, we began an observational
program of IFS of powerful HzRGs at $z\sim$2-3 using VIMOS on the VLT, and
PMAS/PPAK on the 3.5m telescope at the Calar Alto Observatory
(Villar-Mart{\'{\i}}n et al. 2006; Villar-Mart{\'{\i}}n et al. 2007). The main
goal of this programme is to characterize the morphology, kinematics and
ionization conditions of the extended ionized gas surrounding these objects.

In this paper, we present our results based on PPAK observations of two
powerful HzRGs, 4C40.36 ($z=$2.27) and 4C48.48 ($z=$2.34).  Both were selected
from the 4C ultrasteep spectrum compendium (e.g. Chambers et al. 1996a and
reference therein), and are known to have spatialy extended, low surface
brightness emission-line nebulae (Chambers et al. 1996b).  Longslit
spectroscopy for these two sources has been presented by Vernet et al. (2001)
and Villar-Mart{\'{\i}}n et al. (2003).

The layout of this article is as follows: In section \ref{data} we give
details of the observational strategy and data reduction. In section \ref{ana}
we describe the analysis of the data, including the analysis of the two
dimensional distribution of intensity of the different emission lines
(Sec. \ref{ana1}), and the line fitting (Sec. \ref{ana2}). In section
\ref{res} we present the different results from the analysis, describing the
the properties of the nuclear spectra (Sec. \ref{res1}), the morphology of the
emission lines (Sec. \ref{res2}), the ionization conditions at different
locations (Sec. \ref{res3}), and the gas kinematics (Sec. \ref{res4}). In
section \ref{dis} we discuss the results, and summarise the conclusions of
this paper.
 
\section{Observations and data reduction}
\label{data}

The observations were carried out between May the 31st and June the 4th 2005, at
the 3.5m telescope of the Calar Alto observatory, using the Potsdam
MultiAperture Spectrograph (PMAS: \cite{roth05}) in the PPAK mode
(\cite{marc04}; \cite{kelz06}). The atmospheric conditions were stable, with
clear nights, but were not photometric. The seeing was variable, ranging between
0.7$\arcsec$ and 1.2$\arcsec$. 

The PPAK fibre bundle consists of 382 fibres of 2.7'' diameter each (see Fig.5
in Kelz et al.  2006). Of these, 331 fibres (the science fibres) are
concentrated in a single hexagonal bundle covering a field-of-view of
72''$\times$64'', with a filling factor of $\sim$65\%. The sky is sampled by
36 additional fibres, distributed in 6 bundles of 6 fibres each, located
following a circular distribution at $\sim$90'' of the center and at the edges
of the central hexagon. The sky-fibres are distributed among the science ones
in the pseudo-slit, in order to have a good characterization of the sky. The
remaining 15 fibres are used for calibration purposes, as described
below. Cross-talk between adjacent fibres is estimated in less than a 5\% when
using a pure aperture extraction (see below). The distribution of the fibres
in the pseudo-slit does not follow any regular distribution in comparison with
their distribution in the focal plane of the telescope (Kelz et al. 2006),
which minimizes evenmore the effect of the cross-talk. Two different gratings
with different resolutions, wavelength coverage and exposure times were
used. Table \ref{tab:log} gives a log of the observations, including the date,
the grating, the wavelength range covered, the instrumental resolution (FWHM
of the sky-emission lines), the wavelength sampling at the detector, the
target, and the exposure time (number of exposures $\times$ length of
individual exposures). The FWHM of the instrumental profile corresponds to 424
km s$^{-1}$ for the V600 grating, and 672 km s$^{-1}$ for the V300 grating, at
the redshifted wavelength of Ly$\alpha$.

\begin{table}
\caption{Summary of the observations}
\label{tab:log}      % is used to refer this table in the text
\centering                          % used for centering table
\begin{tabular}{lllrrll}        % centered columns (4 columns)
\hline\hline                 % inserts double horizontal lines
Date & Grating & Wavelength & FWHM$^*$& Sampling & Object & Exposure\\
     &         & range (\AA)& (\AA) & (\AA/pix) &        & Time (sec)     \\     
31/05/05 & V600 & 3700-5350 &  3.9 & 1.5 & 4C40.36 & 5x1800 \\
01/06/05 & V600 & 3700-5350 &  3.9 & 1.5 & 4C40.36 & 5x1800 \\
02/06/05 & V300 & 3700-7100 &  7.8 & 3.2 & 4C40.36 & 4x1800 \\
03/06/05 & V600 & 3700-5350 &  3.9 & 1.5 & 4C40.36 & 3x1800 \\
04/06/05 & V600 & 3700-5350 &  3.9 & 1.5 & 4C40.36 & 3x1800 \\
\hline
01/06/05 & V600 & 3700-5350 &  3.9 & 1.5 & 4C48.48 & 6x1800 \\
02/06/05 & V300 & 3700-7100 &  7.8 & 3.2 & 4C48.48 & 5x1800 \\
03/06/05 & V600 & 3700-5350 &  3.9 & 1.5 & 4C48.48 & 5x1800 \\
04/06/05 & V600 & 3700-5350 &  3.9 & 1.5 & 4C48.48 & 3x1800 \\
\\
\hline                                   %inserts single line
\end{tabular}

(*) Errors are too low to be quoted. 

\end{table}

Data reduction was performed using R3D (\cite{sanc06}), in combination with
IRAF packages (\cite{iraf})\footnote{IRAF is distributed by the National
Optical Astronomy Observatories, which are operated by the Association of
Universities for Research in Astronomy, Inc., under cooperative agreement with
the National Science Foundation.} and E3D (\cite{sanc04}). The reduction
consists of the standard steps for fibre-based integral-field spectroscopy. A
master bias frame was created by averaging all the bias frames observed during
the night and was subtracted from the science frames.

The individual exposures from each night (see Table \ref{tab:log}) were then
combined using IRAF routines, clipping any probable cosmic rays. The location
of the spectra in the CCD was determined using a continuum illuminated
exposure taken before the science exposures. Each spectrum was extracted from
the science frames by coadding the flux within an aperture of 5 pixels around
this location along the cross-dispersion axis for each pixel in the dispersion
axis, and was stored in a row-staked-spectrum file RSS
(\cite{sanc04}). Wavelength calibration was performed using a HeHgCd lamp
exposure obtained from archive data, and corrected for distortions using ThAr
exposures obtained simultaneously to the science exposures through the
calibration fibres (indicated above).  Differences in the fibre-to-fibre
transmission throughput were corrected by comparing the wavelength-calibrated
RSS science frames with the corresponding continuum illuminated ones. The sky
emission was determined by using the spectra obtained throughout the
sky-fibres, interpolating to recover the sky-spectrum at the location of any
other fibre as described in S\'anchez (2006), and it was then subtracted from
the science spectra. Each night we obtained spectra of the spectrophotometric
standard star Hz44 that we used to perform the flux calibration. The galactic
extinction is low towards both objects (E(B-V)=0.062 and E(B-V)=0.089 mags,
for 4C40.36 and 4C48.48 respectively), and hence no correction for galactic
extinction was applied to the data. Any uncertainties introduced by not
applying the extinction correction are expected to be quite substantially
smaller than those introduced by our flux calibration. The large size of the
fibres and the low air-mass of the object during the observations minimize the
effects of the differential atmospheric refraction. We checked its possible
effect by inspecting the location of field stars across the field-of-view at
each wavelength, finding no appreciable displacement. Thus, we did not perform
a correction for this effect.

 \begin{figure*}
   \centering \centering \resizebox{\hsize}{!}
   {\includegraphics[width=8cm,angle=-90]{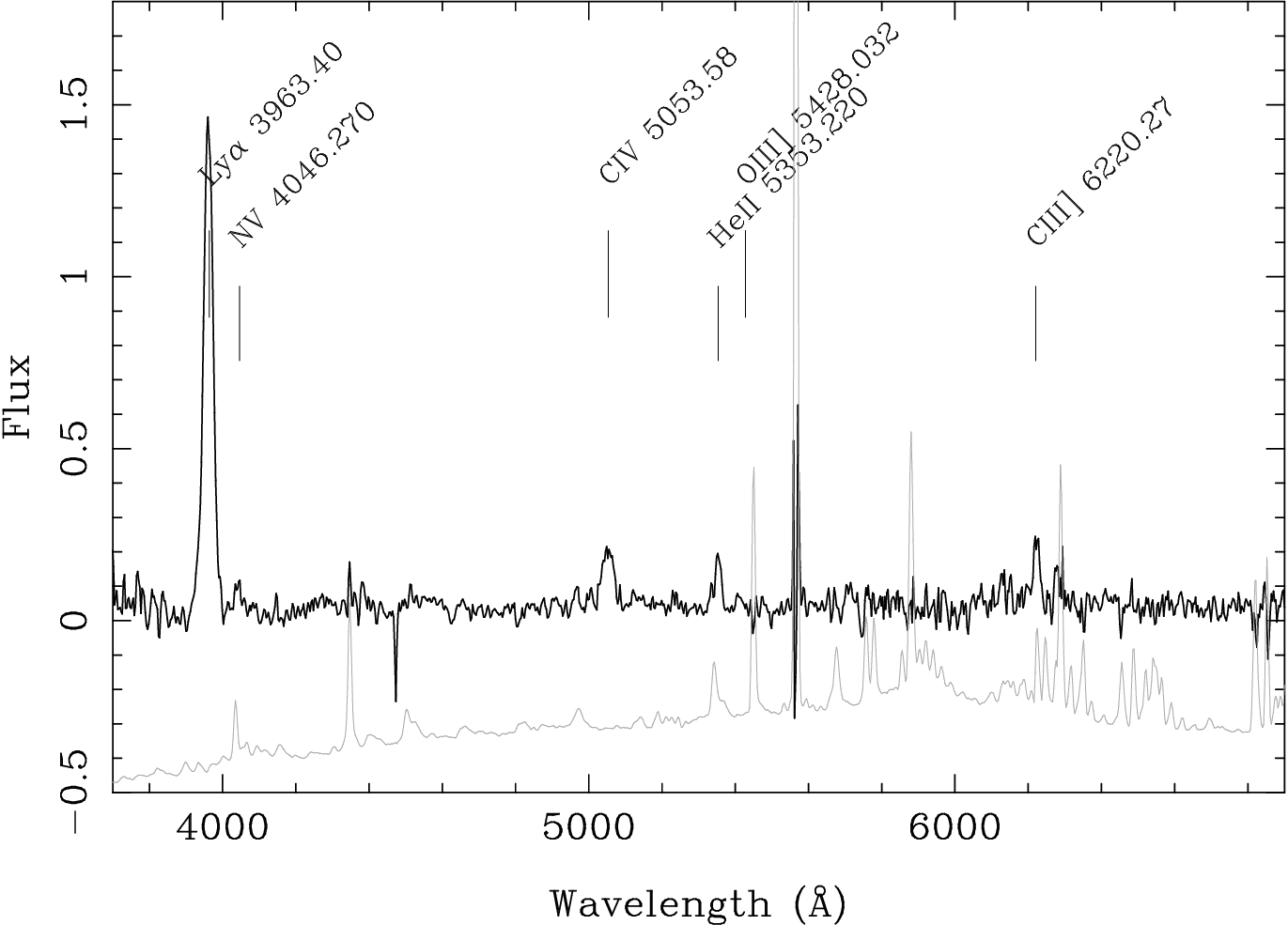}
    \includegraphics[width=8cm,angle=-90]{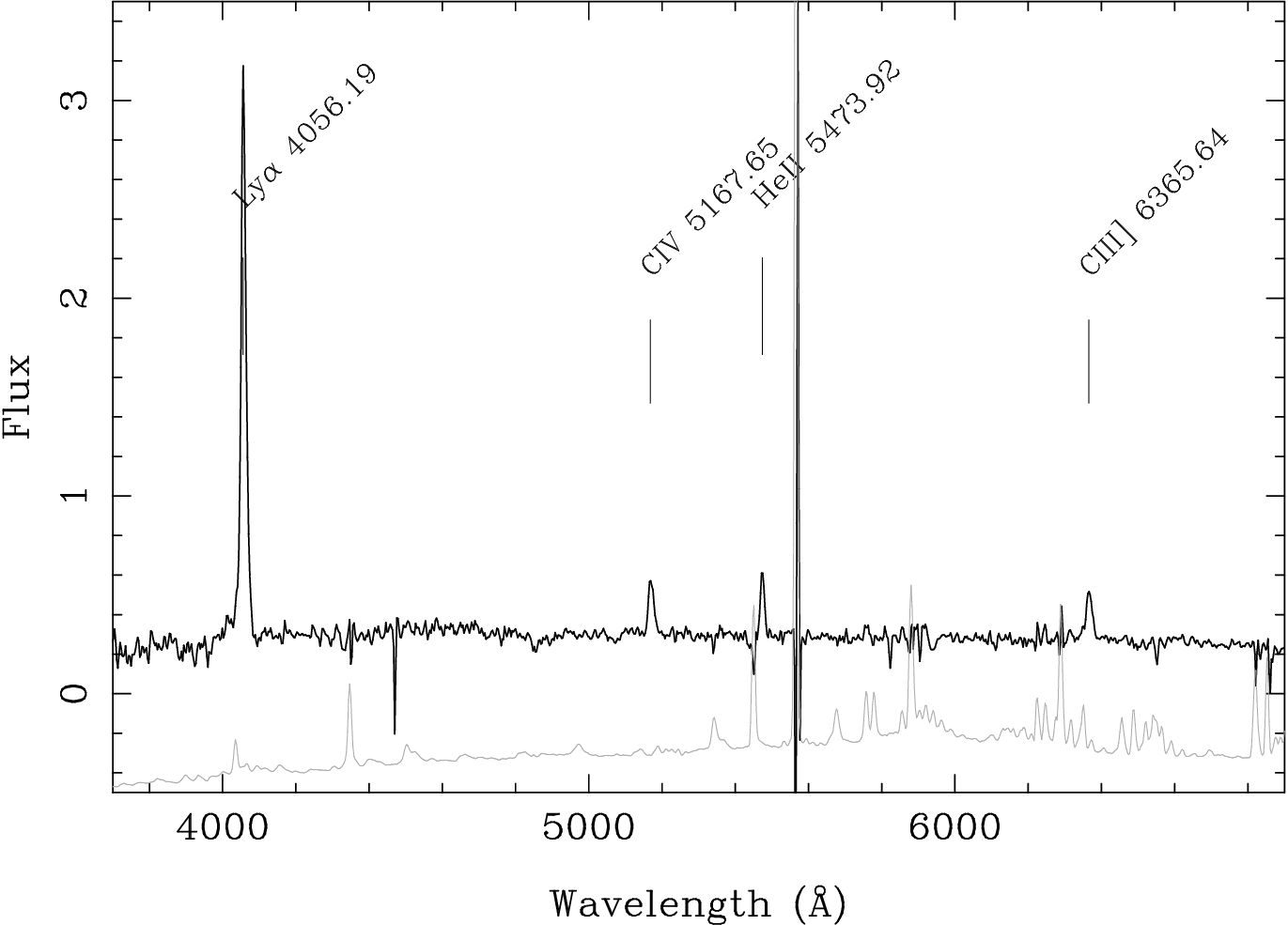}}
 \caption{Spectra extracted from the V300 IFS data for both objects, 4C40.36
 (left panel) and 4C48.48 (right panel), corresponding to the fibre with the
 peak emission in Ly$\alpha$ in units of 10$^{-16}$ erg s$^{-1}$ cm$^{-2}$
 arcsec$^{-2}$ \AA$^{-1}$ (black solid line). The identification of the
 detected emission lines and their observed wavelength are presented.  For
 comparison purposes a scaled version of the night sky emission spectrum has
 been included in both figures (grey solid line). Strong residuals from an
 imperfect sky subtraction are appreciated in the spectra, in particular at
 $\sim$5577\AA. Some emission lines, like HeII in the case of 4C 40.36 may be
 affected by the imperfect subtraction of adjacent night sky emission lines.}
 \label{fig:id}       % Give a unique label
 \end{figure*}

Once reduced, the different exposures taken on different nights were combined to produce a
final frame for each grating setting. The final combined exposures times were
28800s for 4C40.36 with the V600 grating and 7200s with the V300 one, and
25200s for 4C48.48 with the V600 grating and 9000s with the V300 one.  We
estimate the depth of the combined exposures as follows.  First, we measuring the 
standard deviation of the background in a wavelength region free of emission.  
Then, we calculate the flux of an emission line with a FWHM similar to the instrumental 
profile and an average flux per pixel of 3 times the background standard deviation. The resulting
3$\sigma$ detection limit for an emission line ranges between $\sim$2-5 \funitsS for the four
final datacubes. 

\begin{figure*}
   \centering \centering \resizebox{\hsize}{!}
   {\includegraphics[width=\hsize,angle=-90]{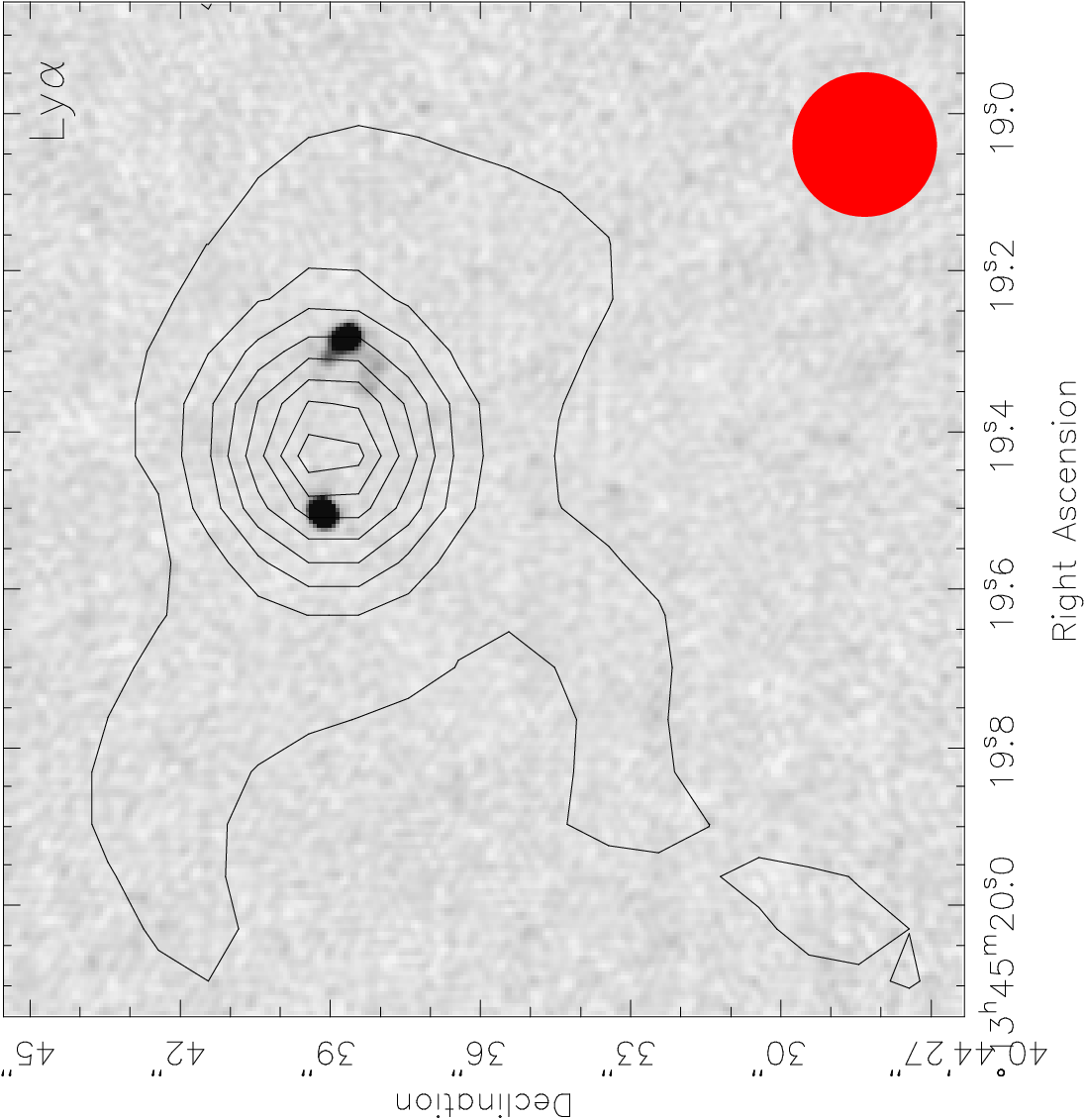}
    \includegraphics[width=\hsize,angle=-90]{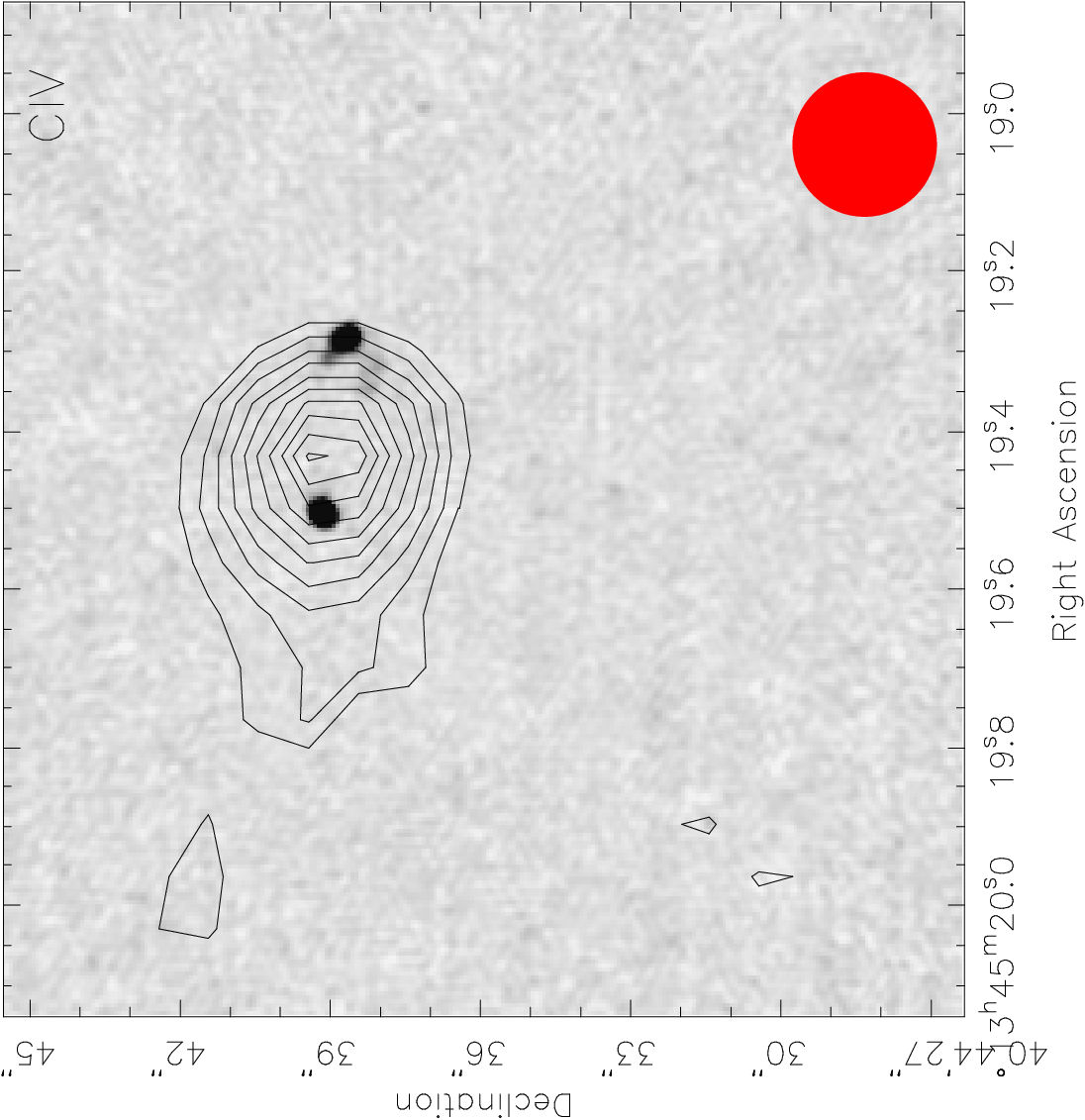}
    \includegraphics[width=\hsize,angle=-90]{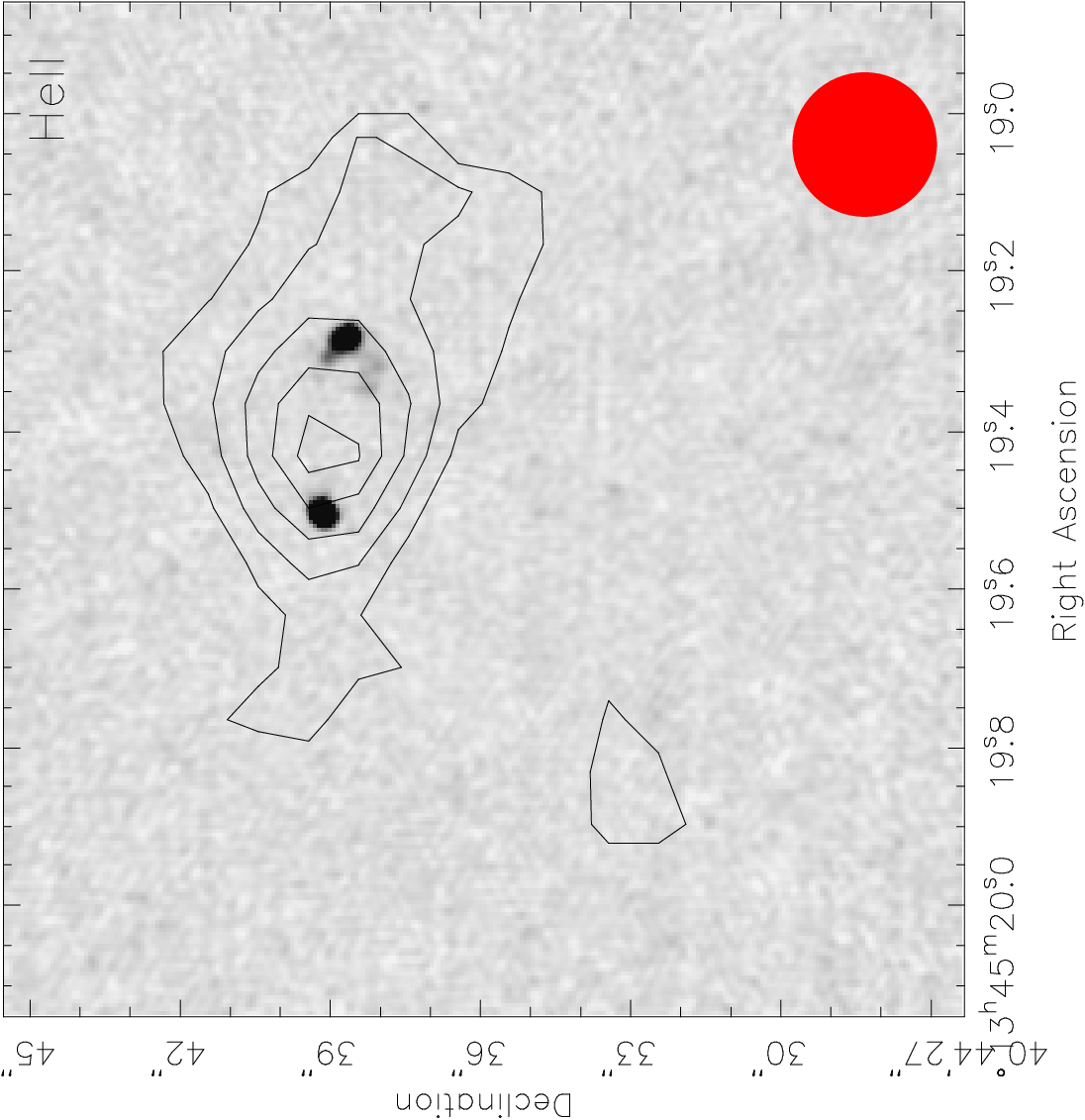}}
   \centering \centering \resizebox{\hsize}{!}
   {\includegraphics[width=\hsize,angle=-90]{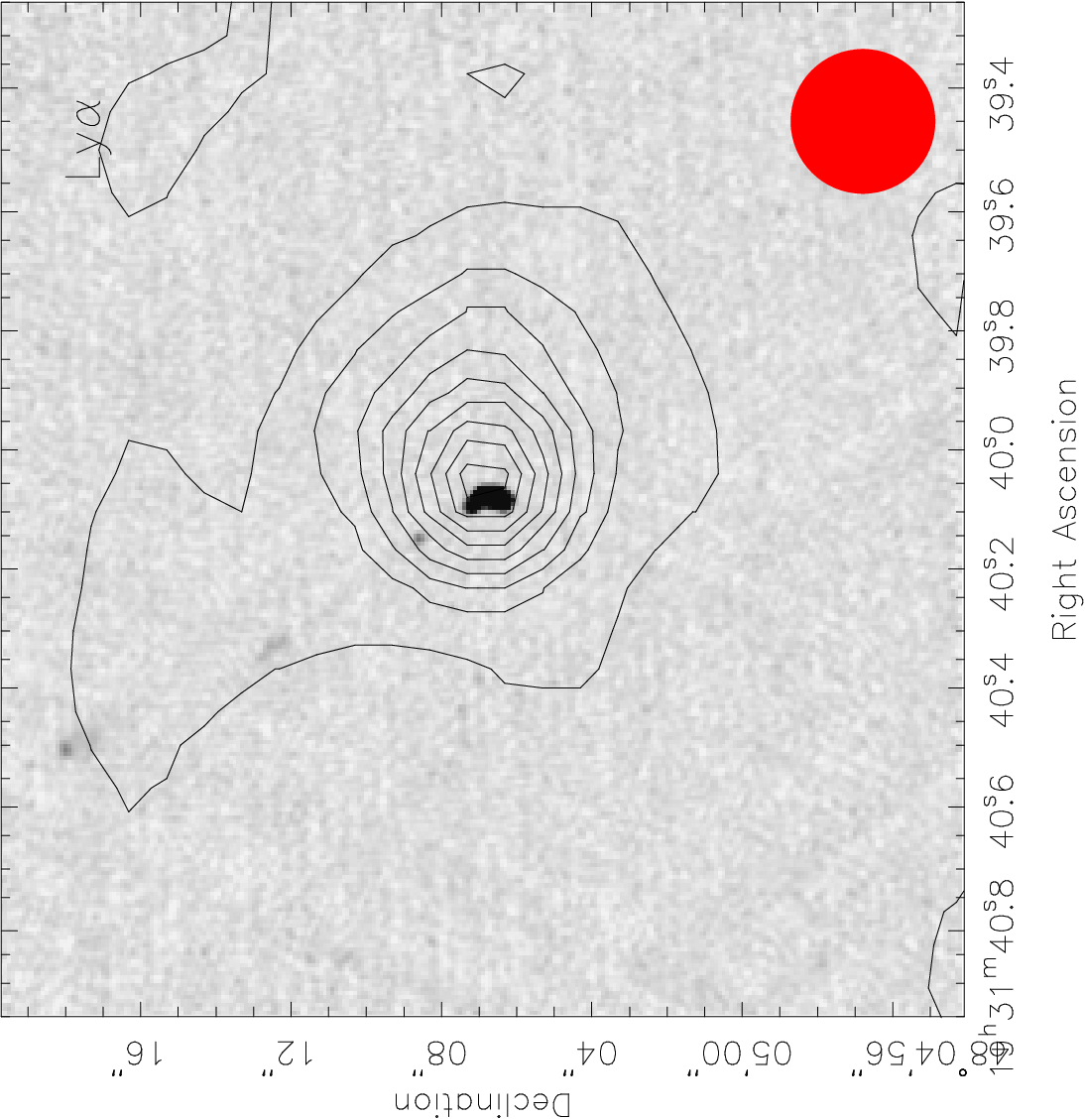}
    \includegraphics[width=\hsize,angle=-90]{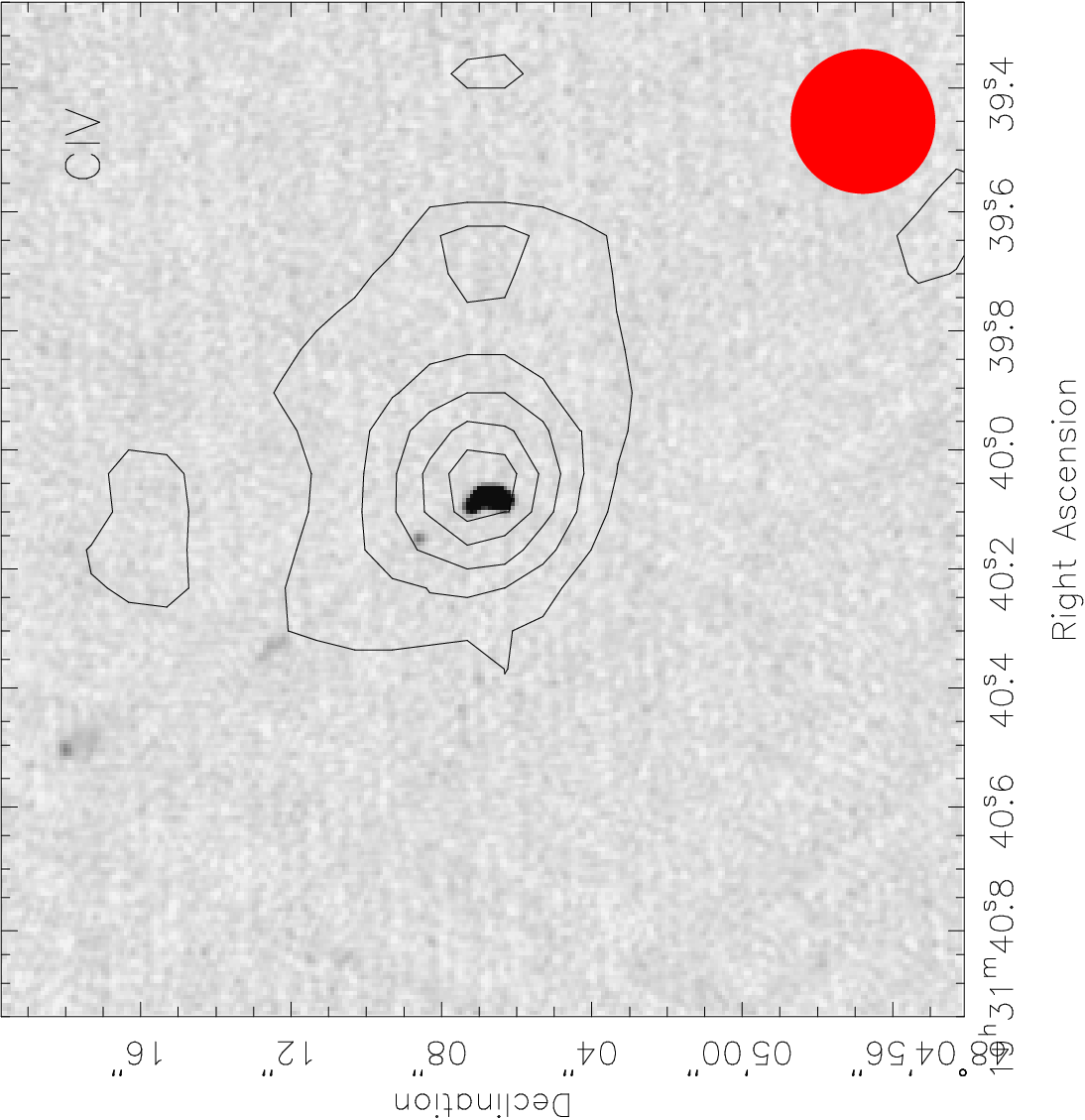}
    \includegraphics[width=\hsize,angle=-90]{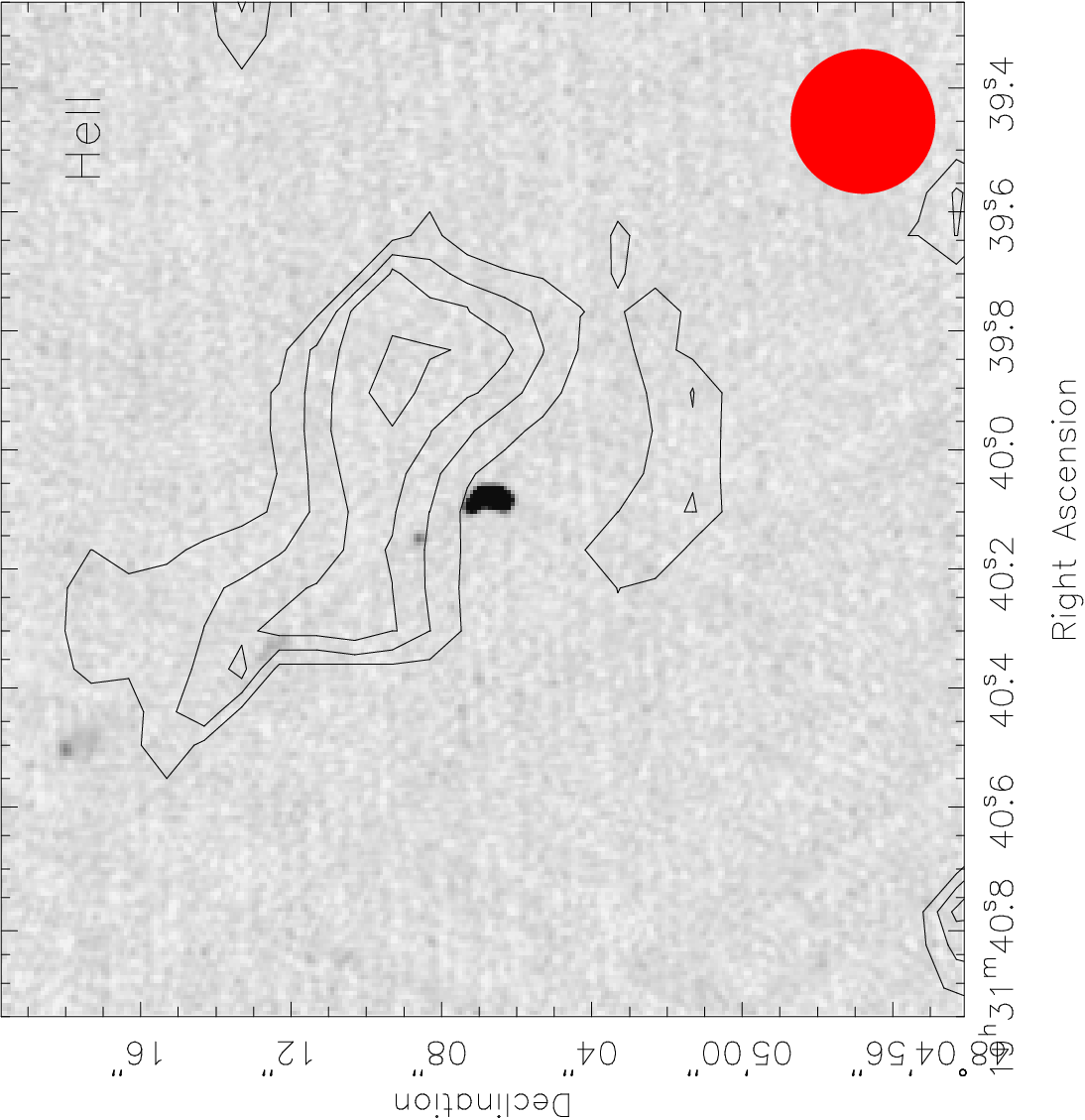}}
 \caption{{\it top-left panel:} Contour plot of a narrow-band image of
 Ly$\alpha$ extracted from the interpolated IFS datacube of the radio galaxy
 4C40.36, together with a grayscale image of the radio map at 5 GHz, obtained
 at the VLA with A configuration (Carilli et al. 1997). Contours start a 1.2
 \funitsS (about $\sim$1$\sigma$ the detection limit), with a separation
 between consecutive contours of 6 \funitsS . {\it top-central panel:} Similar
 contour plot of a narrow-band image of CIV. Contours start a 0.6 \funitsS,
 which a separation between consecutive contours of 0.6 \funitsS. {\it
 top-right panel:} Similar contour plot of a narrow-band image of HeII, with
 similar contour levels. {\it bottom-left panel:} Contour plot of a
narrow-band image of Ly$\alpha$ extracted from the interpolated IFS datacube
 of the radio galaxy 4C48.48, together with a grayscale image of the radio map
 radio map at 5 GHz, obtained at the VLA with A configuration (Carilli et
 al. 1997). Contours start a 1.2 \funitsS, with a separation between
 consecutive contours of 6 \funitsS.  {\it bottom-central panel:} Similar
 contour plot of a narrow-band image of CIV.  Contours start a 0.6 \funitsS,
 with a separation between consecutive contours of 1.2 \funitsS. {\it
 bottom-right panel:} Similar contour plot of a narrow-band image of
 HeII. Contours start a 0.6 \funitsS, with a separation between consecutive
 contours of 3.5 \funitsS. In all the panels the red solid circle indicates
 the original size of the PPAK fibres.}
 \label{fig:slice_rad}       % Give a unique label
 \end{figure*}

\section{Data analysis}
\label{ana}

The wavelength range of our V300 data includes the redshifted Ly$\alpha$
(1216\AA), NV (1240\AA), CIV (1549\AA), HeII (1640\AA), OIII] (1663\AA) and
CIII] (1909\AA) lines.  In the case of the data using the V600 grating, only
Ly$\alpha$ (1216\AA), NV (1240\AA) and CIV (1549\AA) fell within the spectral
coverage.  We extracted the spectrum of the fibre corresponding to the peak
intensity of Ly$\alpha$ by using E3D, for each object and grating (ie., a
2.7$\arcsec$ aperture spectrum). Figure \ref{fig:id} shows these extracted
spectra for each object corresponding to the V300 grating data, together with
a label for each of the previously listed emission lines detected in the
spectra and their corresponding observed wavelength. Based on these
wavelengths the estimated redshifts were $z\sim2.261$ for 4C40.36, and
$z\sim2.336$ for 4C48.48, similar to the objects' nominal redshifts.

\subsection{Emission line images and spatial registration}
\label{ana1}

Prior to perform any further analysis the IFS data were spatially registered
with astrometric calibrated images, in order to place correctly the detected
structures in the sky. For this, we extract narrow-band images at the
redshifted wavelength of each of the detected emission lines and the adjacent
continuum, blueshifted and redshifted with respect to the emission line. The
width of all the narrow-band images were fixed to 116\AA, which basically
corresponds to the typical width of a standard narrow-band filter
(FWHM$\sim$80-100\AA). It is important to note here that we are considering to
wavelengths in the observer-frame. Once extracted the narrow-band images, the
adjacent continua images were combined and subtracted to the emission line
ones, obtaining, for each line, an emission line image free of
continuum. Since the typical FWHM of these lines is much narrower than the
selected width for the narrow-band images it is expected that the noise
increases somewhat after continuum subtration. However, using this width the
obtained emission line images can be directly compared with similar published
narrow-band images. On the other hand, selecting this width we ensure that all
the flux from the emission lines, even the wings, are included in the
narrow-band images. We checked the results creating images with narrower bands
($\sim$50\AA), and we found no significant differences in the results. In
addition a broad-band image comprising all the wavelength range of our
V300-grating data was extracted from the datacubes ($\sim$3700-7100\AA). This
image is dominated by the continuum emission of the sources within the
field-of-view (appart from the HzRGs, in which the Ly$\alpha$ still
dominates).

To register the IFS data we first compare the Ly$\alpha$ emission line images
with those ones published by Chambers et al. (1996). For 4C40.36 they
published a registered U-band image, which samples Ly$\alpha$ at the redshift
of the object. The peak of the Ly$\alpha$ emission is coincident with that of
the UV continuum emission in this object, as shown in the available Keck
spectra (e.g., Villar-Mart{\'{\i}}n et al. 2006, Humphrey 2005). We register the peak
of the Ly$\alpha$ emission image to the coordinates of the peak emission of
the Chambers et al. (1996) U-band image. For 4C48.48 Chambers et al. (1996)
published a registered Ly$\alpha$ narrow-band image of $\sim$60\AA\
bandpass. The Ly$\alpha$ emission shows a well defined peak in this image.
We used the coordinates of this peak to register our Ly$\alpha$ image.  Once
registered the Ly$\alpha$ images the rest of the narrow-band images and IFS
data were registered using this astrometric solution.  Taking into account the
original size of the PPAK fibres, and the errors in the astrometry of the
Chambers et al. (1996) images, we expect an accuracy not worse than
$\sim$0.5$\arcsec$ in the image registration.

We perform an independent check of the performed spatial registration. The
coordinates of the field stars within the field-of-view of the IFS data,
derived from the broad-band continuum image described before, where compared
with those coordinates derived from the corresponding Digitized Sky Survey
(DSS) POSS2 blue-band images. We found 5 stars in the field-of-view of the
4C40.36 IFS data and 9 more in the field-of-view of the 4C48.48 one. The
coordinates of these stars match within an standard-deviation of
$\sim$0.3$\arcsec$, which we consider the typical uncertainty of our spatial
registration.  

Once registered, the line emission maps can be compared with the emission in
other wavelength ranges. Figure \ref{fig:slice_rad} shows, for each object, a
counter-plot of the Ly$\alpha$, CIV and HeII emission map, together with a
radio map at 5 GHz, obtained at the VLA with A configuration (Carilli et
al. 1997).  The remaining emission lines were not included in this comparison
because they were not detected (NV in the case of 4C48.48) or not detected as spatially
extended (CIII] for both objects, and NV in the case of 4C40.36).  Note that
the displayed images have been spatially interpolated to a regular grid using
E3D, with a final sampling of 1$\arcsec$/pix.

\subsection{velocity, FWHM and line ratio maps}
\label{ana2}

Since it is the brightest and most extended of the emission lines within the
spectral range of our data, we used Ly$\alpha$ as a means to extract
information on the kinematics of the warm ionized gas associated with the two
HzRGs.  In each fibre where we detect Ly$\alpha$, the line was fitted with a
single Gaussian function, plus a low order polynomial function to describe the
continuum, using FIT3D (\cite{sanc06b}).  The central wavelength, the FWHM,
and the integrated intensity were free parameters in the fitting process.  The
V300 and V600 datasets were analysed separately due to their different
spectral resolutions.  All values of velocity dispersion were corrected for
the instrumental profile as listed in Table \ref{tab:log}.  The results were
then spatially interpolated using E3D to create regular gridded images, with a
final sampling of 1$\arcsec$/pix.

For both the velocity and velocity dispersion maps, we adopt the values at the
position of the Ly$\alpha$ flux peak as our fiducial zero point.  As a final
step, we co-averaged the V300 and V600 datasets for each of the two HzRGs.
(We note that for each of the two objects, the V300 and V600 datasets are very
similar.)  The velocity and FWHM maps for 4C+40.36 and 4C+48.48 are shown in
\ref{fig:kin}.

\begin{figure}
%   \centering \centering \resizebox{13cm}{!}
%   {\includegraphics[width=6cm,angle=-90]{rat_LyA_CIV_4C40.ps}
%    \includegraphics[width=6cm,angle=-90]{rat_LyA_HeII_4C40.ps}}
   \centering \centering \resizebox{7.5cm}{!}
   {\includegraphics[width=7cm,angle=-90]{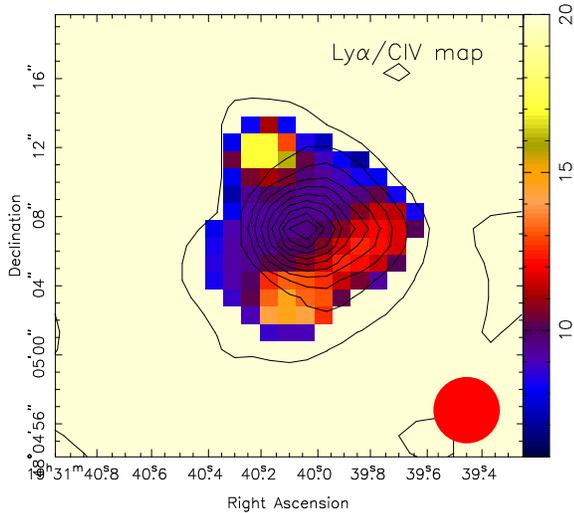}}
 \caption{
%{\it top-left:} Colour coded image of the spatial distribution of the
% Ly$\alpha$/CIV emission line ratio for 4C40.36, together with a contour plot
% of the Ly$\alpha$ emission line intensity, similar to the one shown in the
% corresponding panel of Fig. \ref{fig:kin}. {\it top-right:} Similar plot to
% the one shown in the left panel but for the Ly$\alpha$/HeII emission line
% ratio.
 Colour coded image of the spatial distribution of the
 Ly$\alpha$/CIV emission line ratio for 4C48.48, together with a contour plot
 of the Ly$\alpha$ emission line intensity, similar to the one shown in the
 corresponding panel of Fig. \ref{fig:kin}.  Only where Ly$\alpha$ and CIV are 
both detected to we plot the ratios.  The solid red circle indicates the size 
of the PPAK fibres, prior to resampling the data. }
 \label{fig:rat}       % Give a unique label
 \end{figure}

In the case of 4C+48.48, we produced Ly$\alpha$/CIV line ratio maps by simply
dividing the interpolated Ly$\alpha$ map by those of CIV, respectively.  This
ratio map is shown in figure \ref{fig:rat}.  Other line ratios, such as
Ly$\alpha$/HeII, CIV/HeII or CIV/CIII], have lower signal to noise in the
spatially extended regions or are strongly affected by inaccuracies in the
continuum subtraction, and hence they are not shown.  In the case of 4C+40.36,
the line ratio maps were rather noisy, and thus are also not shown here.

\section{Results and discussion}
\label{res} 

\subsection{The nuclear spectrum}
\label{res1}

Table 2 shows the results from fitting the emission lines in the fibre wherein
the line emission is brightest, which we assume includes the nucleus of the
galaxy.  The kinematic properties derived from the emission in this fibre are
generally consistent with those reported by Villar-Mart{\'{\i}}n et
al. (2003).  Table 3 shows emission line ratios composed from the Ly$\alpha$,
NV, CIV, HeII and CIII] lines in this fibre.  For comparison, we also show the
line ratios measured from Keck II long-slit spectra (Vernet et al. 2001;
Humphrey 2005), for which the slit was aligned with the radio axis.  Vernet et
al. (2001) used relatively large extraction apertures of 4.7\arcsec $\times$
1.0\arcsec and 6.2\arcsec $\times$ 1.0\arcsec for 4C+40.36 and 4C+48.48,
respectively.  Humphrey (2005), analysing the same spectra as Vernet et
al. (2001), but used extraction apertures of 2.1\arcsec $\times$ 1.0\arcsec
for both objects.

From Table 3, it can be seen that several of the line ratios differ
significantly between the different apertures.  In the case of 4C+40.36, the
NV/HeII ratio in our central fibre (and in the 2.5\arcsec $\times$ 1.0\arcsec
aperture of Humphrey 2005) is almost twice that measured in the larger
aperture of Vernet et al. (2001).  For 4C+48.48, the NV/HeII, CIV/HeII,
CIV/CIII] and CIII]/HeII ratios are between $\sim$1.5-2.4 times lower in our
central fibre (and also in the 2.5\arcsec $\times$ 1.0\arcsec aperture of
Humphrey 2005) than in the relatively large aperture of Vernet et al. (2001).
Comparing these differences against ionization models (e.g. Humphrey et
al. 2008a), we consider that the two most plausible explanations are (a) a
spatial gradient in gas metallicity, with higher metallicities closer to the
nucleus, for 4C+40.36, or with the oposite trend, for 4C+48.48 (e.g. Overzier
et al. 2001) or (b) a spatial gradient in the excitation level of the gas
(e.g. Humphrey et al. 2007), with higher excitation nearer the nucleus.  More
emission lines, such as NIV] 1749 or the forbidden optical lines, would be
needed to break the degeneracy between these two possibilities (e.g. Humphrey
et al.  2008a).  We can, however, conclude that within the extended
nebulosities of both 4C+40.36 and 4C+48.48 the ionized gas have a significant
range in physical conditions, as has been found for other HzRGs (e.g. Humphrey
et al. 2007).

\begin{table*}
\caption{Results of the fitting procedure for the peaks emission spectrum}
\label{tab:fit_cen}      % is used to refer this table in the text
\centering                          % used for centering table
\begin{tabular}{lllrrrr}        % centered columns (4 columns)
\hline\hline                 % inserts double horizontal lines
Object  & grating & Line Id. & Wavelength & Flux & FWHM  & FWHM\\
        &         & $^*$     & (\AA)      & $^{**}$  & (\AA)$^{***}$ & \vel \\
\hline                                   %inserts single line
4C40.36 & V300  & Ly$\alpha$          & 3961.03 & 45.83$\pm$0.23 & 28.8 & 2179.7$\pm$557.3\\
        &       & NV                  & 4043.84 &  2.45$\pm$0.23 & 28.8 & 2135.1$\pm$545.9\\
        &       & CIV                 & 5052.75 &  6.58$\pm$0.20 & 35.6 & 2112.2$\pm$540.1\\
        &       & HeII                & 5352.34 &  3.60$\pm$0.20 & 35.6 & 1994.0$\pm$510.0\\
        &       & CIII]$\lambda$1909  & 6222.77 &  3.46$\pm$0.20 & 16.0 &  770.8$\pm$196.9\\
\hline 
        & V600  & Ly$\alpha$          & 3967.39 & 62.11$\pm$0.85 & 27.0 & 2040.2$\pm$228.3\\
        &       & NV                  & 4050.35 &  2.67$\pm$0.85 & 27.0 & 1998.4$\pm$223.6\\
        &       & CIV                 & 5056.13 &  9.45$\pm$0.33 & 21.4 & 1268.9$\pm$142.0\\
\hline\hline
4C48.48 & V300  & Ly$\alpha$          & 4057.26 & 46.66$\pm$0.51 & 14.0 & 1034.5$\pm$274.9\\
        &       & Ly$\alpha$ $1$      & 4058.39 & 37.47$\pm$0.42 & 18.4 & 1359.2$\pm$361.1\\
        &       & Ly$\alpha$ $2$      & 4055.83 & 11.36$\pm$0.28 &  3.9 &  288.2$\pm$\ 76.6\\
        &       & CIV                 & 5170.31 &  5.16$\pm$0.12 & 15.6 &  904.5$\pm$240.3\\
        &       & CIV $1$             & 5172.25 &  4.13$\pm$0.11 & 15.0 &  869.4$\pm$231.0\\
        &       & CIV $2$             & 5165.23 &  1.00$\pm$0.08 &  3.5 & 203.1$\pm$\ 54.0\\
        &       & HeII                & 5474.06 &  4.89$\pm$0.21 & 10.6 &  580.5$\pm$154.2\\
        &       & CIII]$\lambda$1909  & 6366.45 &  4.92$\pm$0.19 & 17.9 &  842.9$\pm$224.0\\
\hline
4C48.48 & V600  & Ly$\alpha$          & 4060.72 &104.29$\pm$1.21 & 13.0 & 959.7$\pm$\ 43.2\\    
        &       & Ly$\alpha$ $1$      & 4063.55 & 51.31$\pm$0.99 & 16.0 &1180.4$\pm$\ 53.2\\    
        &       & Ly$\alpha$ $2$      & 4059.41 & 52.60$\pm$0.78 &  9.5 & 701.6$\pm$\ 31.6\\    
        &       & CIV                 & 5172.39 &  9.30$\pm$0.23 & 16.5 & 956.3$\pm$\ 43.1\\
        &       & CIV $1$             & 5175.85 &  6.16$\pm$0.21 & 16.5 & 955.7$\pm$\ 43.1\\
        &       & CIV $2$             & 5168.53 &  3.17$\pm$0.16 &  8.9 & 516.2$\pm$\ 23.3\\ 
\hline                                   %inserts single line
\end{tabular}

(*) $1$ indicate the broader components and $2$ the narrower one.

(**) {10$^{-16}$ erg s$^{-1}$ cm$^{-2}$ arcsec$^{-2}$}.

(***) After removing the instrumental profile width.

\end{table*}

\begin{table*}
\caption{Line ratios for the total emission and the perturbed and quiescent
  components with it was possible to decouple them.}
\label{tab:rat}      % is used to refer this table in the text
\centering                          % used for centering table
%\begin{tabular}{rrrrrrrrr}        % centered columns (4 columns)
\begin{tabular}{rrlllllll}
\hline\hline                 % inserts double horizontal lines
Object  & dataset & Ly$\alpha$/HeII & Ly$\alpha$/CIV & Ly$\alpha$/NV  &   NV/HeII     & CIV/HeII      & CIV/CIII]    & CIII]/HeII   \\
4C40.36 & V300    &  12.73$\pm$0.23 &  6.97$\pm$0.07 & 18.71$\pm$0.48 & 0.68$\pm$0.03 & 1.83$\pm$0.05 & 1.90$\pm$0.05 & 0.96$\pm$0.03 \\
        & V600    &                 &  6.57$\pm$0.13 & 23.26$\pm$1.45 &               &               &               &               \\
        \hline
        & V01$^*$& 14.7             & 8.4            & 39.2           & 0.37          & 1.8           & 1.8           & 1.0   \\
	& H05$^{**}$ &   11.6          &  7.1           & 19.3          &   0.6        & 1.6           & 2.2           & 0.7 \\     
        \hline
        \hline
4C48.48 & V300    &   9.57$\pm$0.13 &  9.04$\pm$0.12 &   $>$17.9       &  $<$0.38      & 1.06$\pm$0.03 & 1.05$\pm$0.03 & 1.01$\pm$0.03 \\
\multicolumn{2}{r}{Broader}
                  &                 &  9.07$\pm$0.15 &                &               &               &               &               \\
\multicolumn{2}{r}{Narrower}
                  &                 & 11.36$\pm$0.75 &                &               &               &               &               \\
4C48.48 & V600    &                 & 11.21$\pm$0.21 &   $>$65.2       &               &               &               &               \\
\multicolumn{2}{r}{Broader}
                  &                 &  8.33$\pm$0.24 &                &               &               &               &               \\
\multicolumn{2}{r}{Narrower}
                  &                 & 16.59$\pm$0.88 &                &               &               &               &               \\
        \hline                     
	& V01$^*$& 13.2             & 12.6            & 48.3          & 0.27          & 1.0           & 1.7           & 0.6  \\
	& H07$^{**}$ &  11.3           & 8.2            &  12.6          &  0.9         & 1.4           & 2.6           & 0.5 \\
\hline                                   %inserts single line
\end{tabular}

$^{*}$ Vernet et al. (2001)
$^{**}$ Humphrey (2005). 
\end{table*}

\subsection{Emission line morphologies}
\label{res2}

The Ly$\alpha$ emission from 4C+40.36 has a large spatial extent, and is
distributed across $\sim$12$\arcsec$ $\times$ 10$\arcsec$ ($\sim$100 kpc
$\times$ 83 kpc: see figure \ref{fig:slice_rad}).  The CIV and HeII lines from
this object are also spatially extended, and are aligned with the axis of the
radio emission.  As noted by previous investigators (e.g. Chambers et
al. 1996b; Villar-Mart{\'{\i}}n et al. 2003), the Ly$\alpha$, CIV and HeII
emission extends far beyond the radio structure. Their derivation of the
spatial extension of these emission lines was limited by the position and aperture of the
slit in their observations.  Allowing for the low spatial
resolution of our data, the distributions of the lines are broadly consistent
with being within a pair of diametrically opposed `ionization cones' with an
opening angle of $\sim90^{\circ}$, aligned with the radio axis.  We have not
obtained any clear detection of line emission in those regions that are outside
any plausible ionization cone.  From the spatially very extended Ly$\alpha$
emitting regions to the East (A in Figure \ref{fig:slice_rad}) and South-East
(B) of the nucleus, we detect CIV and HeII, respectively.  This shows that the
gas in these regions is ionized, and is not merely a neutral Ly$\alpha$
`mirror' (e.g. Villar-Mart{\'{\i}}n, Binette \& Fosbury 1996).  The most
distant knot of Ly$\alpha$ emission to the South-East of 4C+40.36 (C: at a
radius of $\sim$70 kpc) is also detected in broad band images (Chambers et
al. 1996b), and thus may be associated with a companion galaxy.

Interestingly, the HeII line is brighter on the Western side of 4C+40.36 than
on the Eastern side, while the reverse is true for the CIV line.  This might
be due to a difference in the ionization properties or chemical abundances
between the Western and Eastern emission line regions.  However, we are unable
to discriminate between these two possible explanations, because the CIV/HeII
ratio is sensitive both to the ionization properties and the chemical
abundance of the ionized gas (e.g. Vernet et al. 2001).

In the case of 4C+48.48, the Ly$\alpha$ emission is also very spatially
extended, and is detected across $\sim$14$\arcsec$ $\times$ 16$\arcsec$
($\sim$116 kpc $\times$ 133 kpc).  Its spatial extent parallel to the axis of
the radio source is similar to the extent measured from the slit spectrum of
Villar-Mart{\'{\i}}n et al. (2003).

The Ly$\alpha$, CIV and HeII emission lines from 4C+48.48 show a striking
morphology. They extend several arcseconds towards the North-East, in close
alignment with the radio source.  However, on the other side of the nucleus,
this alignment with the radio source is not present: While the South-Western
radio source undergoes a dramatic $\sim$45 deg bend to the South, the line
emission extends $\sim$5\arcsec ($\sim$40 kpc) from the nucleus towards the
West.  In this object, we note that the CIV and HeII lines have very similar
spatial distributions to that of Ly$\alpha$.  This shows that the giant nebula
is ionized.  Moreover, the presence of HeII emission rules out ionization by
young stars, but is consistent with photoionization by the obscured AGN
(e.g. Vernet et al. 2001). We note here that the two-dimesional extensions of the CIV and
HeII emission lines have been reported for the first time in the current study.

%It is important to note here that the HeII image
%suffers from an imperfect continuum subtraction, due to the proximity of a
%field star, which avoid us to do a more detail analysis of the morphology of
%the Ly$\alpha$/HeII ratio.

\begin{figure*}
   \centering \centering \resizebox{13cm}{!}
   {\includegraphics[width=6cm,angle=-90]{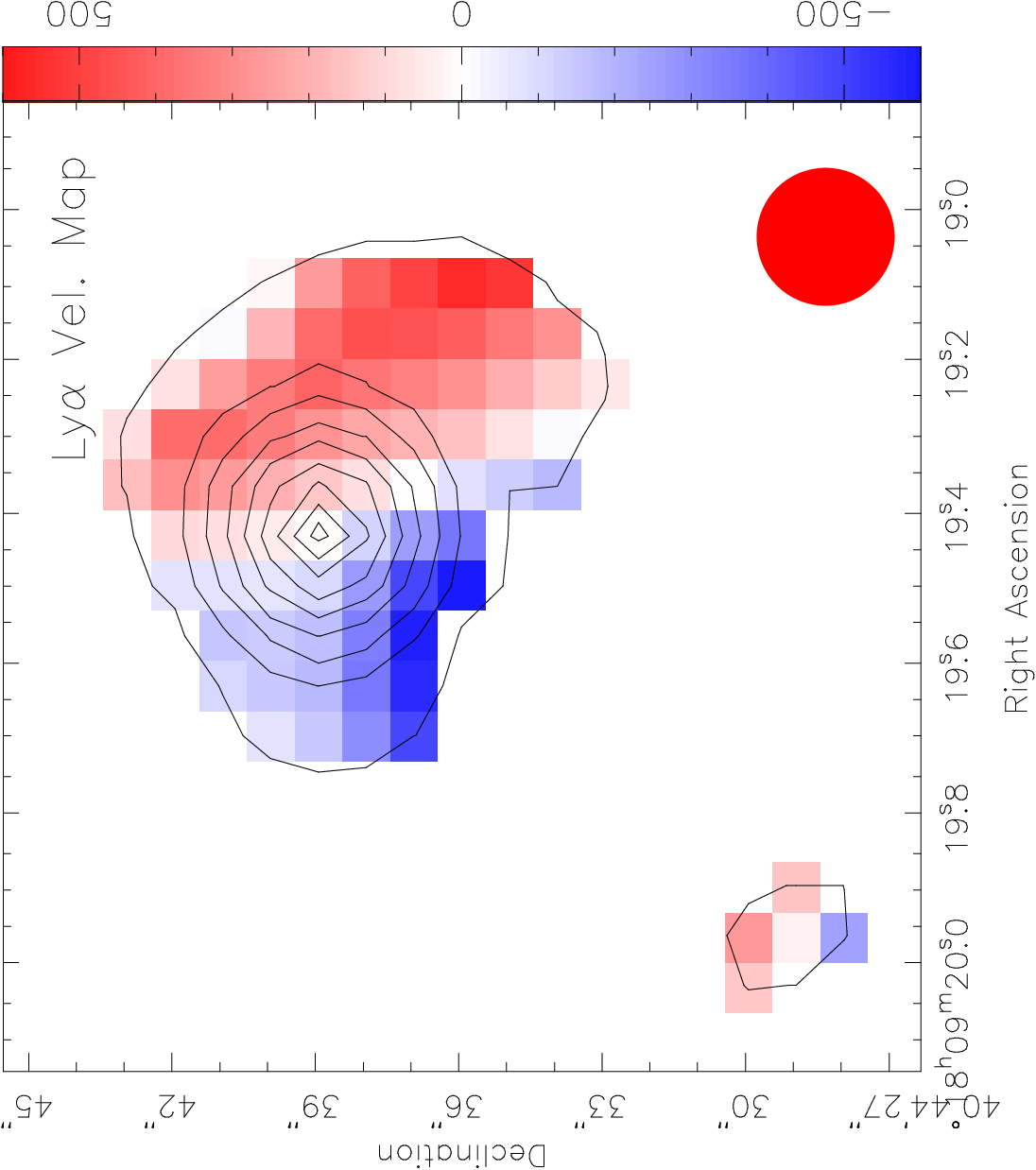}
    \includegraphics[width=6cm,angle=-90]{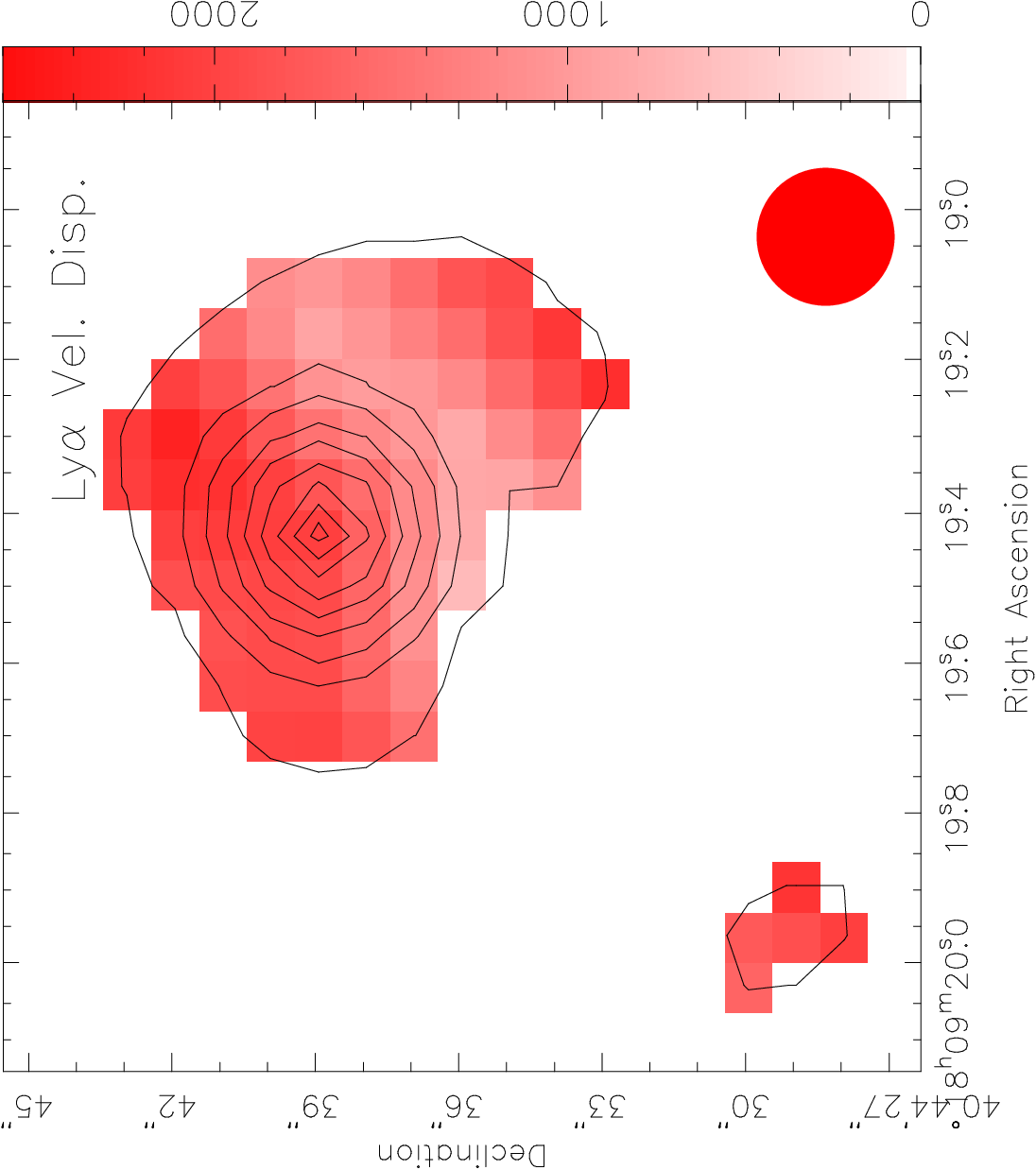}}
   \centering \centering \resizebox{13cm}{!}
   {\includegraphics[width=6cm,angle=-90]{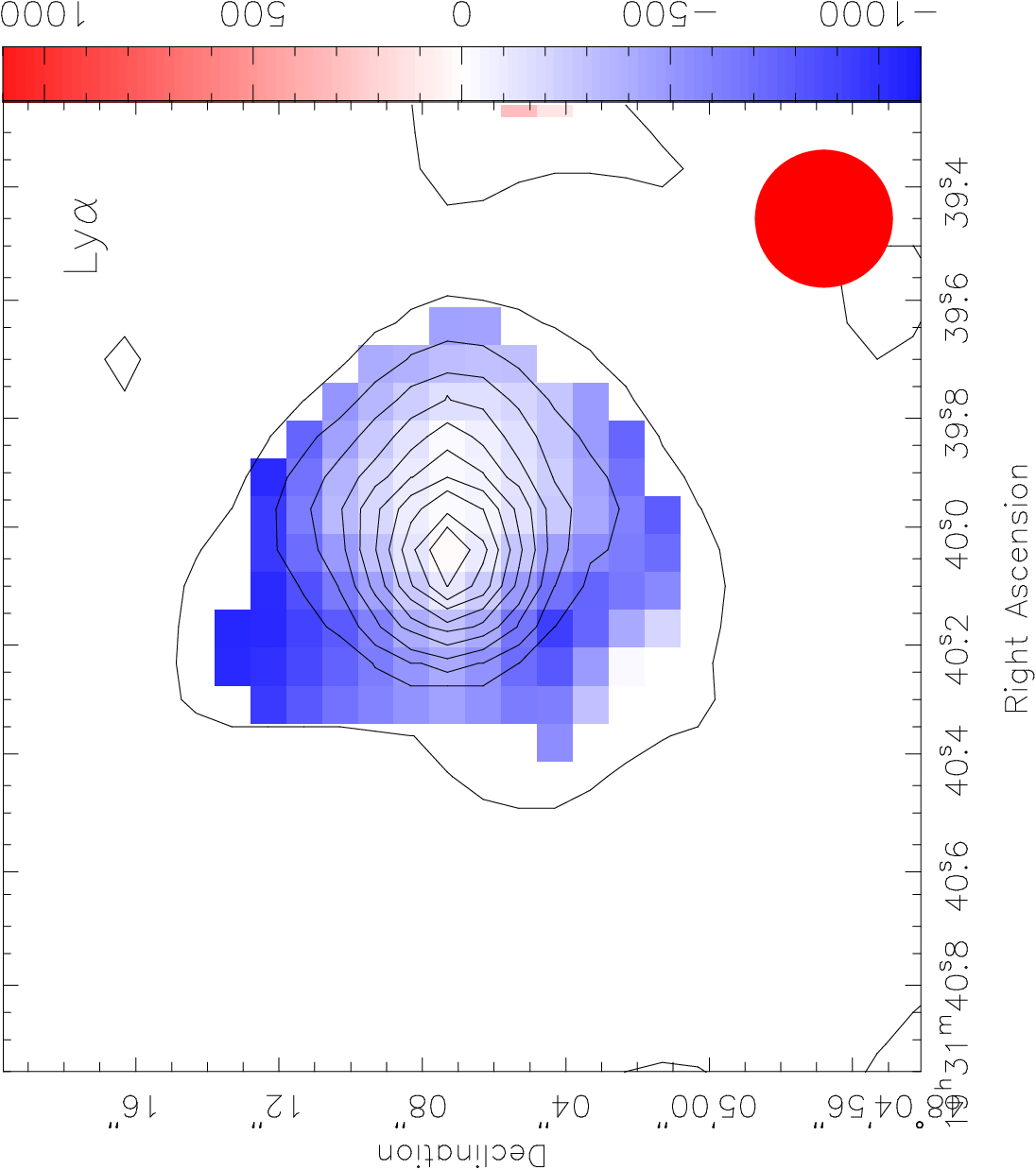}
    \includegraphics[width=6cm,angle=-90]{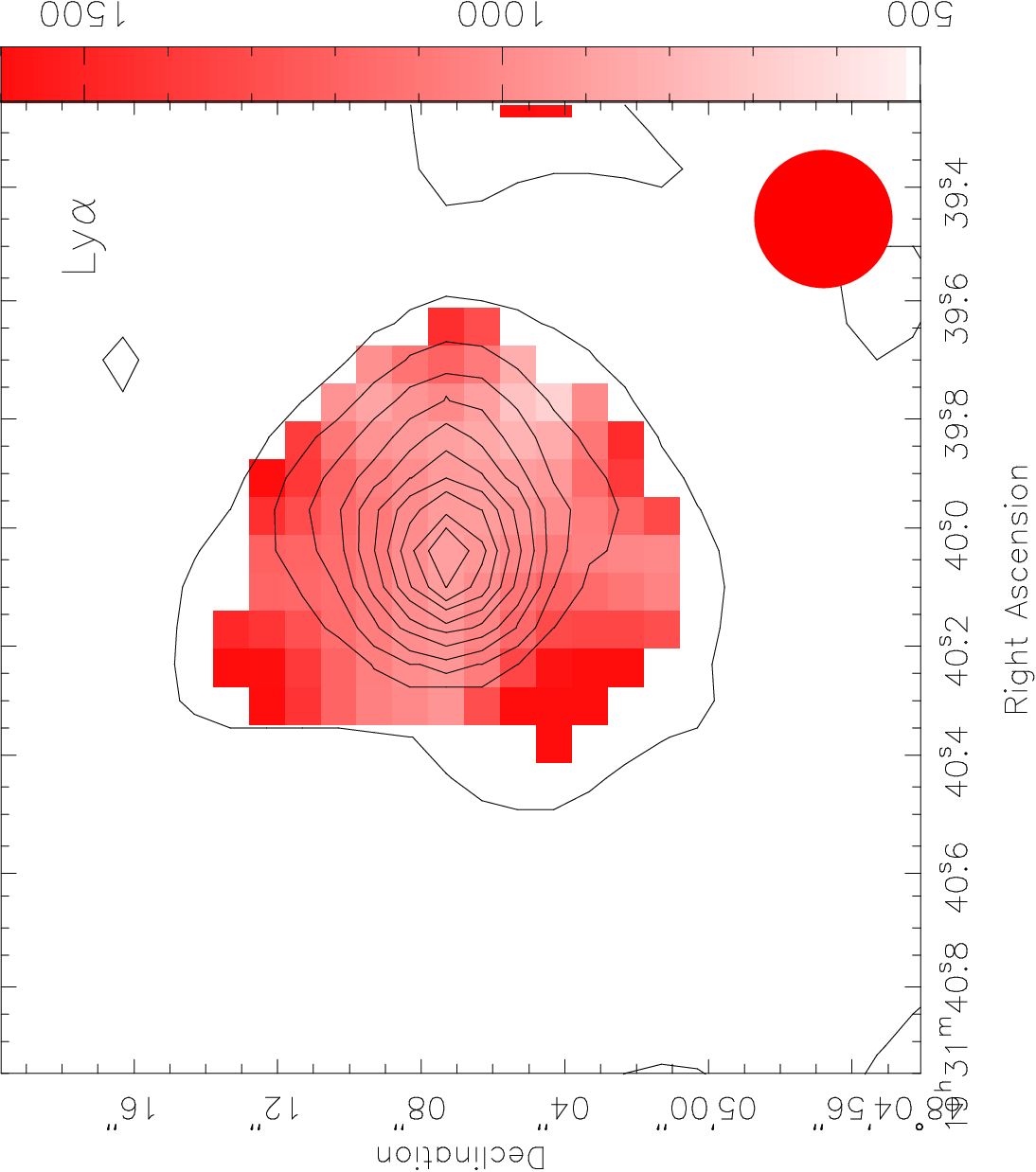}}
 \caption{{\it top-left:} Colour-coded image of the Ly$\alpha$ velocity map of
 4C 40.36 together with a contour plot of the emission line intensity
 recovered by the fitting procedure. Contours start a 1.2 \funitsS, which a
 separation between consecutive contours of 6 \funitsS. Only the velocities
 corresponding to emission line fluxes above 1.2 \funitsS are plotted. {\it
 top-right:} Similar plot for the Ly$\alpha$ velocity dispersion map derived
 by the fitting procedure. {\it bottom-left:} Similar plot to that of top-left
 panel, but for 4C48.48. {\it bottom-right:} Similar plot to that of top-right
 panel, but for 4C48.48. In all the panels the red solid circle indicates the
 original size of the PPAK fibres. }
 \label{fig:kin}       % Give a unique label
 \end{figure*}

\subsection{Ly$\alpha$/CIV map of 4C+48.48}
\label{res3}

Figure \ref{fig:rat} shows the spatial distribution of the Ly$\alpha$/CIV line
ratio.  Since the HeII image is affected by the presence of the nearby field
star, we do not show a Ly$\alpha$/HeII map.  Previous investigations using
long-slit spectroscopy (Humphrey et al. 2007) revealed that line ratios
involving Ly$\alpha$ vary significantly {\it along the radio axis}.  Our
Ly$\alpha$/CIV map confirms their result and, moreover, shows that this ratio
also varies substantially in two spatial dimensions.

The line ratio map reveals a band of relatively low Ly$\alpha$/CIV, running
perpendicular to the radio axis, and passing through the position of maximum
Ly$\alpha$ flux.  In this band, we measure Ly$\alpha$/CIV ratios as low as ~8,
whereas standard photoionization models predict Ly$\alpha$/CIV $\ga$10
(e.g. Villar-Mart{\'{\i}}n et al. 2007).  Thus, we consider these relatively
low observed values of Ly$\alpha$/CIV to be the result of absorption of
Ly$\alpha$, presumably by neutral Hydrogen or interstellar dust.  We suggest
that this band of Ly$\alpha$ absorption might be the observational signature
of an edge-on disc of gas and dust associated with the radio galaxy host
(e.g. Gopal-Krishna \& Wiita 2000).

To the South-West and North-East of the low Ly$\alpha$/CIV band, we measure Ly$\alpha$/CIV ratios of $\sim$11-15, which are consistent 
with the values predicted by standard photoionization models (e.g. Villar-Mart{\'{\i}}n et al. 2007).

\subsection{Emission line kinematics}
\label{res4}

Figure \ref{fig:kin} (top-left panel) shows a colour coded image of the
Ly$\alpha$ velocity map for 4C+40.36, together with a contour plot of the
intensity of this line.  Since we do not detect any stellar absorption lines
from which to determine the systemic velocity, we have simply adopted the
velocity at the position of the Lya emission peak as our fiducial zero point.
It is important to recognise that this velocity may not necessarily correspond
to the true systemic velicity, as discussed by Humphrey et al. (2008)

This velocity map shows a somewhat regular pattern, with a clear gradient from
East to West.  The maximum velocity amplitude is $\sim$1000 km/s, with the gas
to the East approaching and the gas to the West receding, relative to the
fiducial velocity zero-point.  This is consistent with the velocity curves
obtained by Villar-Mart{\'{\i}}n et al. (2003) using a 1\arcsec slit along the
radio axis.  The region of null velocity runs from North-East to South-West,
i.e., at an angle of $\sim$45$^{\circ}$ relative to the axis of the radio and
line emission (see Fig.3).

The relatively ordered nature of the velocity map suggests that the kinematics
of the ionized gas are dominated by ordered motion, i.e., infall, outflow or
rotation.  Superficially, the map is consistent with each of these three
possibilities, and we are unable to definitively distinguish between them.
However, we argue that since the FWHM and velocity amplitude are both roughly
twice what one would expect for gravitational motion around a giant elliptical
galaxy (e.g. Villar-Mart{\'{\i}}n et al. 2002), at least part of the
Ly$\alpha$ nebula associated with 4C+40.36 has been kinematically perturbed,
presumably by the radio source, as suggested by Villar-Mart{\'{\i}}n et
al. (2003).  Moreover, we note that the average FWHM would seem rather large
relative to the velocity amplitude ($\sigma$/v$\sim$1) for a stable rotating
system, even without correcting for inclination.  In addition, we do not
observe the broadening of the FWHM at the null velocity that would be expected
in the case of rotation.

The lower left panel of Fig. 4 shows the colour-coded representation of the
Ly$\alpha$ velocity map of 4C+48.48.  As before, we define our fiducial zero
as the velocity at the peak of the Ly$\alpha$ emission.  The FWHM shows little
spatial variation, with a typical value of 1000 km s$^{-1}$, and with slightly
higher values in the outer regions.  As we also noted for 4C+40.36, we find
that the kinematics of 4C+48.48 are inconsistent with stable rotation: the
FWHM is rather high compared to the amplitude of the velocity shifts, and
there is no clear broadening of the FWHM near the galaxy nucleus or the null
velicity.  Taken at face value, this velocity map is consistent with either
infall or outflow.

\section{Summary}
\label{dis}

In this paper we have presented integral field spectroscopic data-cubes for
two HzRGs, namely 4C+40.36 and 4C+48.48; this study is part of a wider
programme in which we use this technique (Villar-Mart{\'{\i}}n et al. 2006;
Villar-Mart{\'{\i}}n et al. 2007) to obtain information relating to the nature
of the spatially extended gas associated with HzRGs, and their possible
connection with the nuclear and radio-jet acivity.

Although both objects have been observed before using slit-spectroscopy, at
the same wavelength range, we present for the first time IFS data on
them. This techinique allows to study the spatial extension of the emission in
the direction perpendicular to the radio-jet axis, what in many cases
long-slit spectroscopy systematically miss.

For both sources, the Ly$\alpha$ emission is extended across 100 kpc or more,
in agreement with previous imaging and spectroscopic studies (Chambers et
al. 1996b; Villar-Mart{\'{\i}}n et al. 2003).  The CIV and HeII emission lines
are also spatially extended.  While these emission lines are generally aligned
with the axis of the radio emission, we also detect emission far from this
axis.

Our map of the Ly$\alpha$/CIV ratio in 4C+48.48 has revealed a band of low
Ly$\alpha$/CIV running perpendicular to the radio axis, through the assumed
position of the active nucleus.  We suggest that this feature might be the
observational signature of an edge-on disc of neutral gas (e.g. Gopal-Krishna
\& Wiita 2000).

We have argued that when viewed in two spatial dimensions, the kinematic
properties of both 4C+40.36 and 4C+48.48 are inconsistent with stable
rotation.  However, their properties are not inconsistent with infall
(e.g. Humphrey et al. 2007) or outflows (Nesvadba et al. 2006).

\begin{acknowledgements}
 
%  We thank Dr.J.Alves, director of the Calar Alto Observatory, for supporting
%  this work and providing S.F.S\'anchez with the required working time to
%  finish it.
  
  SFS thanks the Spanish Plan Nacional de Astronom\'\i a program
  AYA2005-09413-C02-02, of the Spanish Ministery of Education and Science and
  the Plan Andaluz de Investigaci\'on of Junta de Andaluc\'{\i}a as research
  group FQM322.  
  
  We thank M. Villar-Martin for her fundamental contribution to this project.

 We thank the referee, Dr. L.Binette, who has indicate valuable changes to
  increase the quality of the article.

\end{acknowledgements}

\end{document}